\documentclass[twocolumn,showpacs,preprintnumbers]{revtex4}
\usepackage{amsmath}
\usepackage{amssymb}
\usepackage{graphicx}
\usepackage{dcolumn}
\usepackage{bm}
\usepackage{amssymb}
\usepackage{graphicx}
\usepackage{dcolumn}
\usepackage{bm}

\begin{document}

\title{Effect of magnetic fields on the spin evolution of non-polarized \\
$^{87}$\textbf{Rb} Bose-Einstein condensates }
\author{Zhifeng Chen, Chengguang Bao, Zhibing Li\footnote{%
Corresponding author: stslzb@mail.sysu.edu.cn}}
\affiliation{The State Key Laboratory of Optoelectronic Materials and Technologies \\
School of Physics and Engineering \\
Sun Yat-Sen University, Guangzhou, 510275, P.R. China}

\begin{abstract}
The spin mixing dynamics of spin-1 Bose-Einstein condensates with
zero magnetization and under an external magnetic field is
investigated. The time-dependent solutions are obtained via a
diagonalization of the Hamiltonian, which has a simple form under
the single mode approximation. The features of evolution are
compared in detail with those with the field removed so as to
emphasize the effect of the field, which can induce strong
oscillation in population of atoms in spin component
 $0$ . A new mode of
oscillation characterized by a high frequency and a low frequency,
is found when the field is sufficiently strong.
\end{abstract}

\pacs{03.75.\ Fi, \ 03.65.\ Fd}
\maketitle

\section{Introduction}

The rich physics in the Bose-Einstein condensate of atoms with spin degrees
of freedom has attracted considerable interests in the last decade. \cite%
{ho98,ohmi98,stam98,sten98,law98,goel03,b01,lewe07} Experimentally,
one has been able to prepare spinor condensates of given number of
atoms and given magnetization, and then to observe their evolutions.
The time-dependent population of each spin component has been
measured by the Stern-Gerlach splitting technique. Various modes of
oscillation of the populations have been found, they are sensitive
to the initial states. \cite{pu99,chan04,dien06} Coherent dynamics
has been observed experimentally. \cite{kuwa04,mur06,kron06,kawr07}

When an external magnetic field is applied, the evolution will be
affected. This is a way to control the evolution. Efforts along this
line have been made. \cite%
{pu00,roma04,chan05,grie05,kron05,y06} Most theoretical calculations
on this problem are based on the mean-field theory. In this
approach, the spin dynamics can be modeled as a nonrigid pendulum
which displays a variety of periodic oscillations in magnetic
fields. \cite{zhan05,chan07} The dependence of the oscillation
period on the magnetic field has been confirmed by experiments.
\cite{blac07} In fact, theoretical studies
suggested that more complicate spin dynamic behavior would be possible. \cite%
{more07,dien06}

 In a previous paper of us \cite{luo}, the spin evolution without a magnetic field
 has been studied. Where, instead of using the mean field theory, the time-dependent
 solutions of the Hamiltonian have been obtained under the single mode
 approximation (SMA). Thereby an analytical formula governing the evolution has been derived. The
   present paper is a direct generalization of the previous one, the aim is to clarify the
   effect of magnetic field on the spin evolution. For this purpose, the Hamiltonian arising
    from the SMA has been exactly diagonalized. Thereby the
     time-dependent populations of spin-0 component starting from various initial states can be obtained.
      We have focused on the condensates with zero magnetization. Since the solutions of our Hamiltonian
       are exact, both short-term and long-term behaviors can be predicted by the theory. In the follows
       the basic approach is given in Section II. The features of spin-evolution without a magnetic field
is briefly reviewed in Section III. Those under a magnetic field are
presented in Section IV. In Section V, the effects of atom number
and the confinement potential are discussed. Finally in Section VI,
a summary is given.

\section{Hamiltonian and its solution}

Let $U_{ij}=(c_{0}+c_{2}\mathbf{F}_{i}\cdot \mathbf{F}_{j})\delta (\mathbf{r}%
_{i}-\mathbf{r}_{j})$ be the spin-dependent interaction among $N$
spin-1 atoms. Under the SMA, all the atoms have the same spatial
normalized wave function $\phi (\mathbf{r)}$. Accordingly, the
Hamiltonian for the spin evolution under a magnetic field $B$ with
direction lying along the Z-axis reads
\begin{equation}
H=g\hat{S}^{2}+q\sum_{i}F_{zi}^{2}  \label{e1}
\end{equation}
where $g=\frac{1}{2}c_{2}\int d\mathbf{r}|\phi (\mathbf{r})|^{4}$. \ The
second term arises from the quadratic Zeeman effect, where $%
q=(E_{+}+E_{-}-2E_{o})/2$ is the energy difference of the three Zeeman
levels, which is\ related to $B$ as $q=\mu _{B}^{2}B^{2}/(\hbar
^{2}E_{HFS}), $ and $E_{HFS}$\ is the hyperfine splitting. The linear Zeeman
term does not affect the dynamics due to the conservation of the total
magnetization $M$.

Let a Fock-state be denoted as $|N_{1},N_{0},N_{-1}\rangle $, where
$N_{\mu } $\ is the number of atoms in spin component $\mu $. \
Since $N_{\pm 1}=(N-N_{0}\pm M)/2$,\ due to the conservation of $N$
and $M$ which are initially given, the Fock-state can be simply denoted as $|N_{0}\rangle $ with $N_{0}$%
\ ranged from $N-M$, $N-M-2$, to 0 or 1. \ They form a complete set of bases
for the many particle spin space. The matrix elements of $H$\ read
\begin{eqnarray}
\langle N_{0}^{\prime }|H|N_{0}\rangle &=&[gA_{0}(M,N_{0})+q(N-N_{0})]\delta
_{N_{0},\ N_{0}^{\prime }}  \notag \\
& & +gA_{+}(M,N_{0})\delta _{N_{0},\ N_{0}^{\prime }+2}  \notag \\
& &+gA_{-}(M,N_{0})\delta _{N_{0},\ N_{0}^{\prime }-2}  \label{e2}
\end{eqnarray}
where
\begin{equation}
A_{0}(M,N_{0})=M^{2}+N+N_{0}+2NN_{0}-2N_{0}^{2}  \notag
\end{equation}
\begin{equation}
A_{+}(M,N_{0})=[N_{0}(N_{0}-1)(N+M-N_{0}+2)(N-M-N_{0}+2)]^{1/2}  \notag
\end{equation}
\begin{equation}
A_{-}(M,N_{0})=[(N_{0}+1)(N_{0}+2)(N+M-N_{0})(N-M-N_{0})]^{1/2}  \notag
\end{equation}

After the diagonalization of the Hamiltonian, the eigenenergies $E_{j}$ and
eigenstates $|\phi _{j}\rangle $\ can be obtained, and we have $H|\phi
_{j}\rangle =E_{j}|\phi _{j}\rangle $, $|\phi _{j}\rangle
=\sum_{N_{0}}c_{N_{0}}^{j}\ |N_{0}\rangle $. \ It is emphasized that both $%
E_{j}$ and $\phi _{j}$\ are exact for the Hamiltonian because the space for
the diagonalization is complete.

Let the initial state be a Fock-state with given $N$, $M$, and $%
N_{0}|_{t=0}=I$, and is denoted as $|I\rangle $. \ Then the total spin-state
at time $t$ is
\begin{eqnarray}
\Psi (t)&=&e^{-iHt/\hbar }|I\rangle =\sum_{j}e^{-iE_{j}t/\hbar }|\phi
_{j}\rangle \langle \phi _{j}|I\rangle  \notag \\
&=&\sum_{j}c_{I}^{j}e^{-iE_{j}t/\hbar }|\phi _{j}\rangle  \label{e3}
\end{eqnarray}

It is emphasized that (3) is an exact time-dependent solution of the
Hamiltonian, every detail of the evolution can be thereby extracted.
We are interested in the $t-$dependence of the population of spin-0
component, namely, $\langle \Psi (t)|\overset{\wedge }{N_{0}}%
|\Psi (t)\rangle /N\equiv \mathbf{P}_{I}^{M}(t)$, which can be derived from
(3) as
\begin{equation}
\mathbf{P}_{I}^{M}(t)=\mathbf{B}_{I}^{M}+\mathbf{O}_{I}^{M}(t)  \label{e4}
\end{equation}
where
\begin{equation}
\mathbf{B}_{I}^{M}=\sum_{j}(c_{I}^{j})^{2}%
\sum_{N_{0}}(c_{N_{0}}^{j})^{2}N_{0}/N  \label{e5}
\end{equation}
\begin{equation}
\mathbf{O}_{I}^{M}(t)=2\sum_{j<j^{\prime }}\cos [(E_{j^{\prime
}}-E_{j})t/\hbar ]c_{I}^{j}c_{I}^{j^{\prime
}}\sum_{N_{0}}c_{N_{0}}^{j}c_{N_{0}}^{j^{\prime }}N_{0}/N  \label{e6}
\end{equation}
Eq.(\ref{e4}) implies an oscillation around the background $\mathbf{B}%
_{I}^{M}$.

We consider the condensate of $^{87}$Rb atoms trapped by a harmonic
potential $\frac{1}{2}m\omega ^{2}r^{2}$. In the present paper, $\hbar
\omega ,\ mG$ and $\sec $ are used as units.\ For $^{87}$Rb, $g\simeq
-6.57\times 10^{-5}(\omega /N^{3})^{1/5}$ (evaluated under the Thomas-Fermi
approximation), and $q\simeq 144\pi \times 10^{-6}B^{2}/\omega $. Numerical
results with discussions are given in the follows.

\section{Evolution without magnetic field}

In order to understand the effect of magnetic field $B$, we first
review
the evolution with $B=0$. The evolution is described by the formula of \cite{%
luo}, which is exact for the Hamiltonian (\ref{e1}) with $q=0$. In order to
obtain numerical results from the formula, $N=400$ and $\omega =3000/\sec $
are firstly assumed. Then the effect of $N$ and $\omega $ will be studied.
The evolution has the following features:

(i) $\mathbf{P}_{I}^{M}$ is strictly periodic with a\ period $t_{p}=\pi
/(|g|\omega )$ and symmetric with respect to $t_{p}/2$. $\mathbf{O}%
_{I}^{M}(t)$ is anti-symmetric with respect to $t_{p}/4$. Therefore, the\
study is sufficient to be confined in the duration (0, $t_{p}/4$). Under the
Thomas-Fermi approximation, $t_{p}\approx 1.521\pi (N/\omega
^{2})^{3/5}\times 10^{4}\sec $\ (for $^{87}$Rb). For the given $N$ and $%
\omega $, the above duration is 29.24$\sec $.

(ii) For zero-polarized condensates ($M=0$), all the
$\mathbf{P}_{I}^{0}$ have nearly the same background
$\mathbf{B}_{I}^{0}\approx 1/2$ disregarding $I$ as shown by
eq.(\ref{e9}) of \cite{luo} and by Fig.\ref{CF1}a of the present
paper.

(iii) If $N$\ is even, $I$\ must be even so that $M$ can be zero.\ When $I=0$%
, starting from the initial state that half atoms are spin-up while the
other half are spin-down, $\mathbf{P}_{0}^{0}$ goes up directly to $\mathbf{B}%
_{I}^{0}\approx 1/2$ without oscillation as shown by the black curve
in \ref{CF1}a. From this curve we know that half of the $\mu \neq 0$
atoms become $\mu =0$\ within $0.01t_{p}$, it certainly implies the
occurrence of strong spin-flips $\uparrow +\downarrow
\longleftrightarrow 0+0$. Then it
remains extremely steady in a very long time until $t$ is close to $t_{p}/4$%
, where a round of strong oscillation occurs suddenly.

(iv) When $I=2$, a round of oscillation emerges in the early stage
(red curve). When $I=2k$, $k$ rounds of oscillation emerge (blue
curve). In general, $\mathbf{P}_{I}^{0}$ oscillates in the early
stage but suddenly becomes very steady (this was first found in
\cite{law98}), then the oscillation suddenly recovers. Therefore
there are zones of oscillation (ZOO) and zones of steady evolution
(ZOS). They appear alternately, namely,
ZOO-ZOS-ZOO as shown by the dark cyan curve, and again. In general, when $%
I<N/2$, a larger $I$\ leads to a broader ZOO and accordingly a narrower ZOS.
Incidentally, if $N $\ is odd, $I$\ is odd. When $I=2k+1 <N/2$, the ZOO contains $%
k+1/2$\ rounds of oscillation.

(v) When $I=N/4$, $N/8$ rounds of oscillation are contained in the ZOO (dark
cyan curve), It turns out that, in this occasion, the duration of the ZOO is
just $t_{p}/12$, that of the ZOS\ is also $t_{p}/12.$ Therefore the two ZOO
and the one ZOS of the dark cyan curve all have the same duration $t_{p}/12$%
. Meanwhile, the average frequency of oscillation in the ZOO is equal to $%
3N/(2t_{p})\approx 3.14(N\omega ^{3})^{2/5}\times 10^{-5}$, which $\approx
5.13/\sec $ for our parameters. It can be imagined that, when $N$\ is very
large, the frequency would be very high and the ZOO would appear as a band
with a width.

(vi) When $I$\ is close to $N/2$, the duration of the ZOO becomes very long
(close to $t_{p}/8)$, and the amplitude of oscillation becomes very small
except in a small domain close to $t_{p}/8$ as shown by the green curve in %
\ref{CF1}a. Accordingly, the ZOS in between becomes very narrow and
unsteady.

(vii) When $I=N/2$, the two previous ZOOs transform to ZOSs, while the
previous ZOS becomes to a very narrow ZOO containing only one round of
oscillation situated at $t_{p}/8$ as shown by the top curve in \ref{CF1}a.

Fig.\ref{CF1}a is for the cases with $I\leq N/2$. The cases with $I\geq N/2$
are given in Fig.\ref{CF3}a. It was found that $\mathbf{P}_{I}^{0}$ and $%
\mathbf{P}_{N-I}^{0}$ are one-to-one roughly similar, however they
are greatly different when $t$\ is close to 0 or $t_{p}/4$. The
early stage of evolution is referred to Fig.\ref{CF2}a and
\ref{CF4}a.

\section{Evolution under a magnetic field}

When $B\neq 0$, via the procedure of diagonalization, we obtain the
following numerical results.

(I) Weak field

\begin{figure}[tbp]
\scalebox{1}{\includegraphics{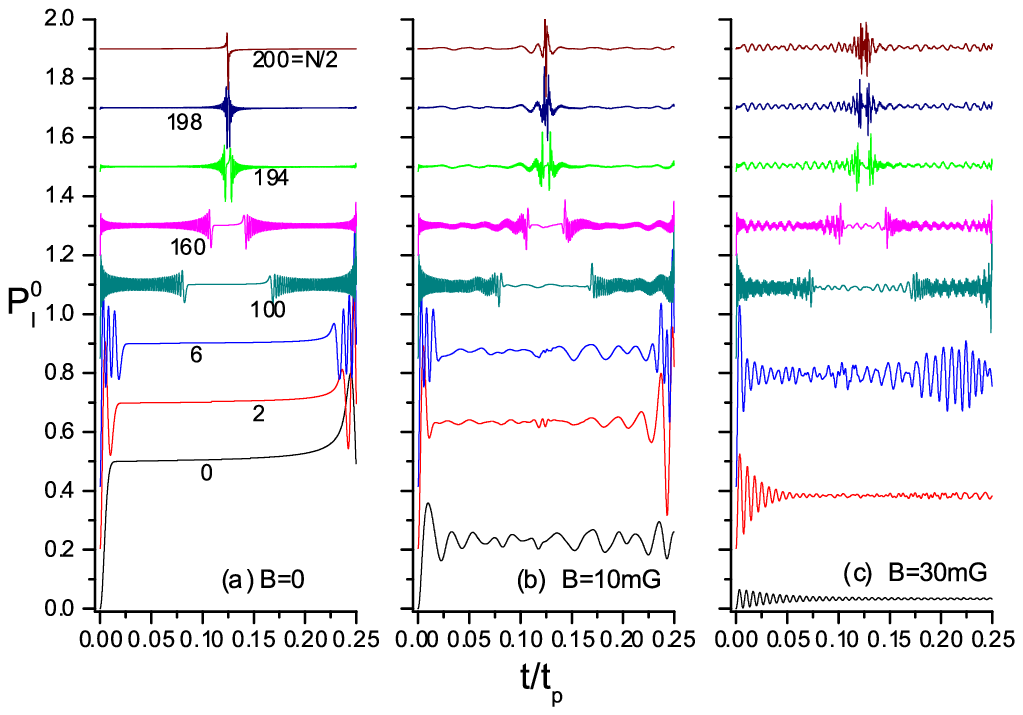}}
\caption{(colored online) Evolution of $\mathbf{P}_{I}^{0}(t)$ with
various $I$\ and $B$. All figures in this paper are for $^{87}$Rb
atoms with $\protect\omega =3000$
and $N=400$ (the only exception of $N$ is Fig.6). Accordingly, the period $%
t_{p}=117$sec, and $t$\ is given from 0 to $t_{p}/4$ (the case with
$t > t_{p}/4$ can be understood from the inherent symmetry).\ $I$ is
from 0 to $N/2$\ marked by the curves. Each curve has been shifted
up by 0.2 more than its lower neighbor. } \label{CF1}
\end{figure}

\begin{figure}[tbp]
\scalebox{1}{\includegraphics{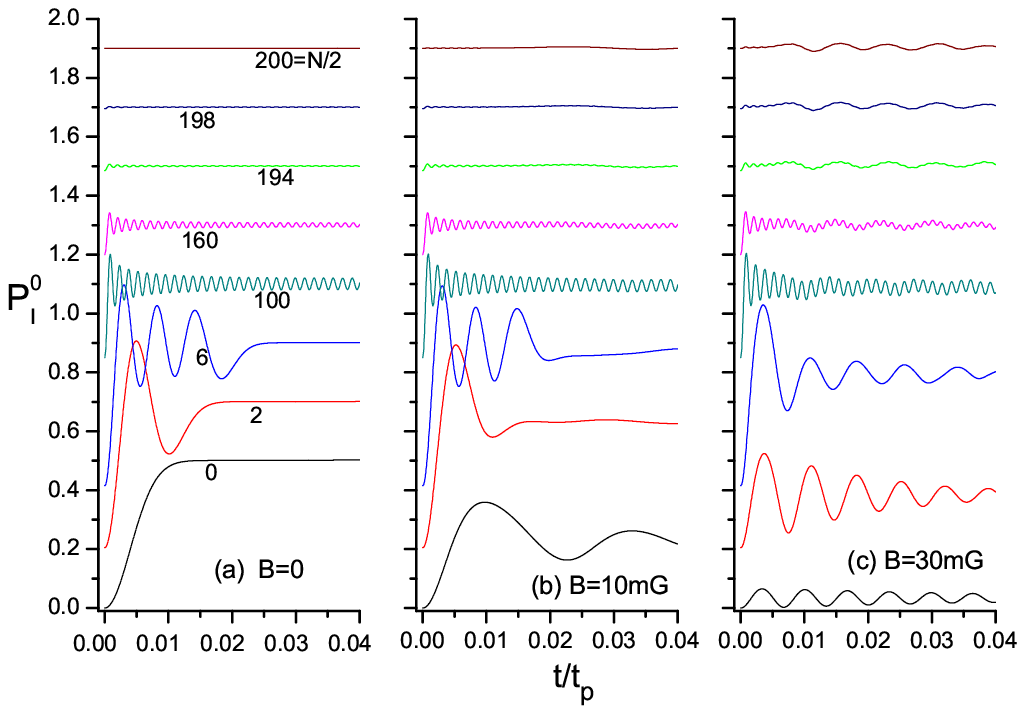}}
\caption{(colored online) The same as Fig.\protect\ref{CF1} but
given only in the early stage. } \label{CF2}
\end{figure}

\begin{figure}[tbp]
\scalebox{1}{\includegraphics{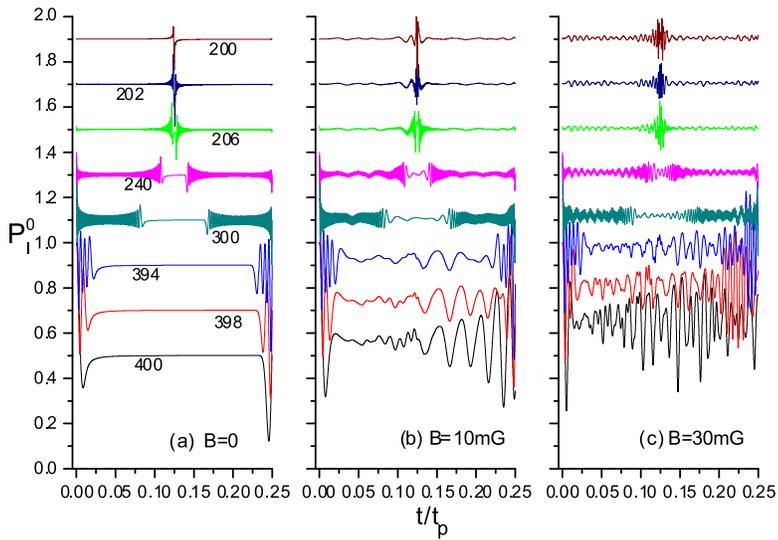}}
\caption{(colored online) The same as Fig.\protect\ref{CF1} but with
$I$\ from $N/2$ to $N$.} \label{CF3}
\end{figure}

\begin{figure}[tbp]
\scalebox{1}{\includegraphics{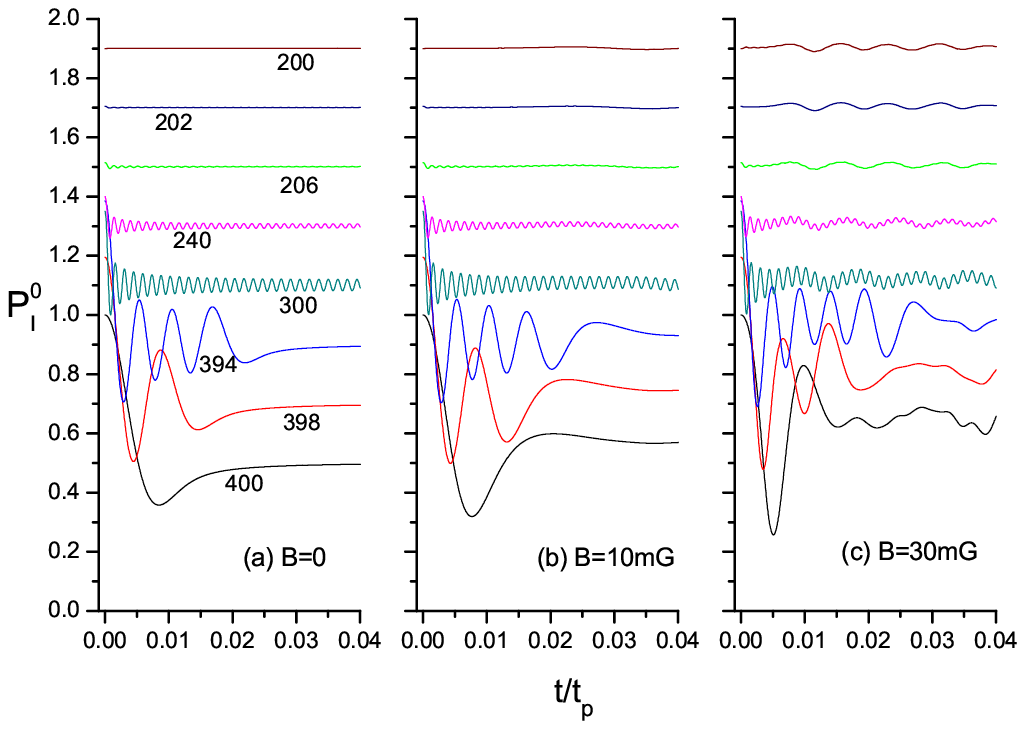}}
\caption{(colored online) The same as Fig.\protect\ref{CF3} but
given only in the early stage. } \label{CF4}
\end{figure}

The evolutions of $\mathbf{P}_{I}^{0}$ are given in the (b) and (c) panels
of Fig.\ref{CF1} to \ref{CF4}, we found

(i) The effect of $B$\ is very sensitive if $I$\ is close to 0 or
$N$.
Otherwise $\mathbf{P}_{I}^{0}$ is less affected. Note that $\mathbf{B}%
_{I}^{0}$ of the lowest three curves in \ref{CF1}a (\ref{CF3}a) are much
higher (lower) than their $I/N$. The effect of $B$ is to pull them down
(push them up) so that $\mathbf{B}_{I}^{0}$ are closer to $I/N$.

(ii) $\mathbf{P}_{0}^{0}$ and $\mathbf{P}_{N}^{0}$ are extremely
sensitive to $B$. This is shown by the black curves in \ref{CF1}b
and \ref{CF3}b. The black curve representing $\mathbf{P}_{0}^{0}$ in
Fig.1a is greatly pulled down by the field when $B=10$ (\ref{CF1}b),
and it becomes nearly a
horizontal line close to zero when $B=30$ (\ref{CF1}c). Thus, for $\mathbf{P}%
_{0}^{0},$ the spin-flips which occurs strongly in the early stage
is severely suppressed by the field.

For $\mathbf{P}_{N}^{0}$ with all the atoms in $\mu =0$ initially,
we know from Fig.\ref{CF3} that 64\% (74\%) of the atoms would be
changed to $\mu \neq 0$ at 0.008$t_{p}$ ( 0.005$t_{p}$) if $B=0$
($B=30$). Thus, instead of being suppressed, the strong spin-flips
in the early stage are accelerated by $B$. Consequently, a strong
oscillation is thereby induced. When $B$=10, the oscillation is
stronger in the duration ($t_{p}/8$,$t_{p}/4$), refer to the black
and red curves of \ref{CF3}b. When $B$=30, the oscillation is
extremely strong through out nearly all the time (black curve of \ref{CF3}%
c), this is a noticeable point and we will return to it.

(iii) When $I$\ is neither close to 0 nor $N$, the effect of $B$\ would be
to shorten the ZOO and enlarge the ZOS if $I<N/2$\ (e.g., the duration of
the ZOO of the dark cyan curve of 1c is $t_{p}/13.2$\ instead of $t_{p}/12$
in \ref{CF1}a), or reversely if $I>N/2$ (e.g., refer to the pink curves in
Fig.\ref{CF3}). Another effect of $B$ is to spoil the stability of the ZOS,
and cause ``irregular" oscillation in both ZOO and ZOS.

(iv) It seems that a new mode of oscillation might be caused by $B$. This is
hinted, say, by the wavy pink curves in \ref{CF3}b, \ref{CF3}c, and the blue
curve in \ref{CF1}c. This point is further studied below.

(v) The strict periodicity and the inherent symmetry exist no more if $B\neq
0$.

(II) Strong field

\begin{figure}[tbp]
\scalebox{1}{\includegraphics{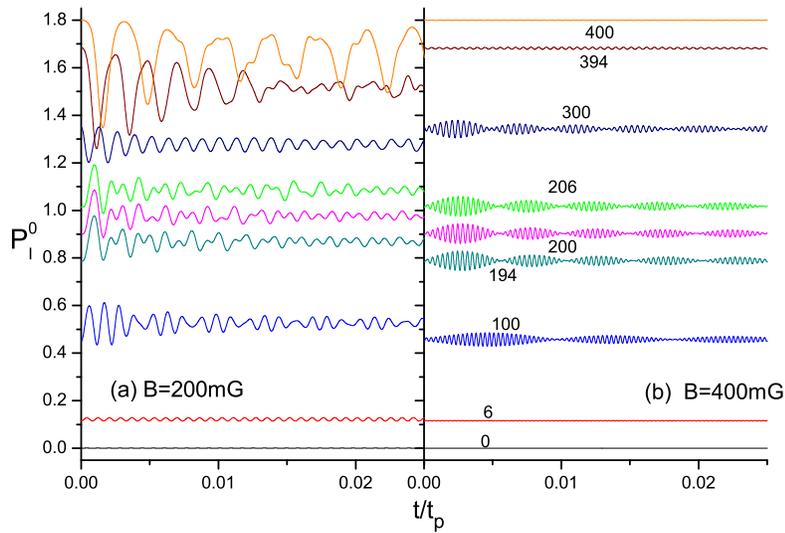}}
\caption{(colored online) Early stage of evolution of $\mathbf{P}_{I}^{0}(t)$ under a strong $B$%
.\ $I$ is from 0 to $N$\ marked by the curves. Each curve has been
shifted up by 0.1 more than its lower neighbor. $t$\ is from 0 to
$t_{p}/40$. } \label{CF5}
\end{figure}

The early stage of evolution of $\mathbf{P}_{I}^{0}$ under a strong
field is plotted in Fig.\ref{CF5} within the duration (0,
$t_{p}/40$). Comparing 5a with Fig.\ref{CF2}c, we found that the
curves with $I$\ close to zero are further suppressed so that
$\mathbf{P}_{I}^{0}\approx $ $I/N$\ together with a negligible
oscillation. Comparing 5a with Fig.\ref{CF4}c, the curves with $I$\
close to $N$ keep their oscillation. The curves with $I$\ neither
very small nor very large are relatively less affected, and they
become more or less similar to each other as shown by the blue,
pink, and navy curves of \ref{CF5}a. For these curves the division
into ZOO and ZOS holds no more.$\ \mathbf{B}_{I}^{0}$ of all the
curves in 5a are close to $I/N$.

When the field is even stronger as in \ref{CF5}b, a new mode of oscillation
emerges, where the new patterns contain a series of pulses, each has an
olivary shape and includes certain rounds of oscillation inside. The number
of round would decrease if $I$ increases.\ The amplitudes of the oscillation
is larger if $I\approx N/2$, and it will be too small to be seen if $%
I\approx 0$ or $N$. If $B$ increases further, the olives would become longer
and thinner, and include more rounds. In the strong $B$\ limit\ the
oscillation is completely suppressed and all $\mathbf{P}_{I}^{0}$ become
just horizontal lines.

When $B$ is sufficiently large, the first term of the Hamiltonian $g\hat{S}%
^{2}$\ can be considered as a perturbation. \ The set of the first order
perturbative solutions of the Hamiltonian is
\begin{equation}
|\Phi _{I}\rangle =|I\rangle +\frac{g}{2q}A_{-}(M,I)|I+2\rangle -\frac{g}{2q}%
A_{+}(M,I)|I-2\rangle  \label{e7}
\end{equation}
With this set, $\Psi (t)$\ (eq.(\ref{e3})) can be calculated analytically.
Accordingly, we have
\begin{eqnarray}
\mathbf{P}_{I}^{M}(t)&\approx& I/N+ \frac{g^{2}}{Nq^{2}} [A_{-}^{2}(M,I)(1-%
\cos \alpha t)  \notag \\
& &-A_{+}^{2}(M,I)(1-\cos \alpha ^{\prime }t)]  \label{e8}
\end{eqnarray}
where $\alpha =[2q-g(4N-8I-6)]\omega $, $\alpha ^{\prime }=\alpha -16g\omega
.$\ Eq.(\ref{e7}) and (\ref{e8}) holds if $|\frac{g}{2q}A_{\pm }(M,I)|<<1.$

When $M=0$\ and $I=0,\ A_{+}(0,0)=0$ and $A_{-}(0,0)=\sqrt{2}N$, thus
\begin{equation}
\mathbf{P}_{0}^{0}(t)\approx 2\frac{Ng^{2}}{q^{2}}(1-\cos \alpha t)
\label{e9}
\end{equation}

When $M=0$\ and $I=N,\ A_{-}(0,N)=0$ and $A_{+}(0,N)=2N$, thus
\begin{equation}
\mathbf{P}_{N}^{0}(t)\approx 1-4\frac{Ng^{2}}{q^{2}}(1-\cos \alpha ^{\prime
}t)  \label{e10}
\end{equation}

Both (\ref{e9}) and (\ref{e10}) imply a small oscillation around a
horizontal line as shown in \ref{CF5}b, where the amplitude of the
black curve is too small to be seen.

When $I$ is neither close to 0 nor $N-M$, the sum of $A_{-}^{2}(M,I)$ and $%
A_{+}^{2}(M,I)$ is much larger than their difference. \ Thus we have
\begin{eqnarray}
\mathbf{P}_{I}^{0}(t)&\approx& I/N+\frac{g^{2}}{Nq^{2}}%
[A_{-}^{2}(0,I)-A_{+}^{2}(0,I)-(A_{-}^{2}(0,I)  \notag \\
& & +A_{+}^{2}(0,I))\sin (\frac{\alpha +\alpha ^{\prime }}{2}t)\sin (\frac{%
\alpha -\alpha ^{\prime }}{2}t)]  \label{e11}
\end{eqnarray}

Since $\alpha +\alpha ^{\prime }$ is much larger than $\alpha -\alpha
^{\prime }$, (\ref{e11}) describes a high frequency oscillation included in
a low frequency oscillation as shown in Fig.\ref{CF5}b. This explains the
origin of the olivary shapes. The factor $\sin (\frac{\alpha -\alpha
^{\prime }}{2}t)$ \ has a period $\pi /(4|g|\omega ).$ Therefore the length
of the olive would tend to $\pi /(8|g|\omega )=t_{p}/8$. Furthermore, $%
\mathbf{P}_{I}^{0}(t)$ would oscillate around $I/N+\frac{g^{2}}{Nq^{2}}%
[A_{-}^{2}(0,I)-A_{+}^{2}(0,I)]$, where the second term is small, as shown
in (\ref{e11}) and in Fig.\ref{CF5}b.

\section{The effect of the confinement potential and the particle number}

The above numerical results are obtained with $\omega =3000$ and $N=400$.
From the formula of our previous paper \cite{luo} we know that the change of $%
\omega $ is simply equivalent to a change of the scale of time, i.e., all the curves of
$\mathbf{P}_{I}^{M}(t)$ would remain unchanged except that the implication of $t_{p}$
is changed with  $%
\omega $ via the relation $t_{p}\propto \omega ^{-6/5}$ . On the
other hand, since $g\propto \omega ^{1/5}$, the decrease of $\omega
$\ will slightly increase the ratio $q/g$, therefore will amplify
slightly the effect of $B$.

\begin{figure}[tbp]
\scalebox{1}{\includegraphics{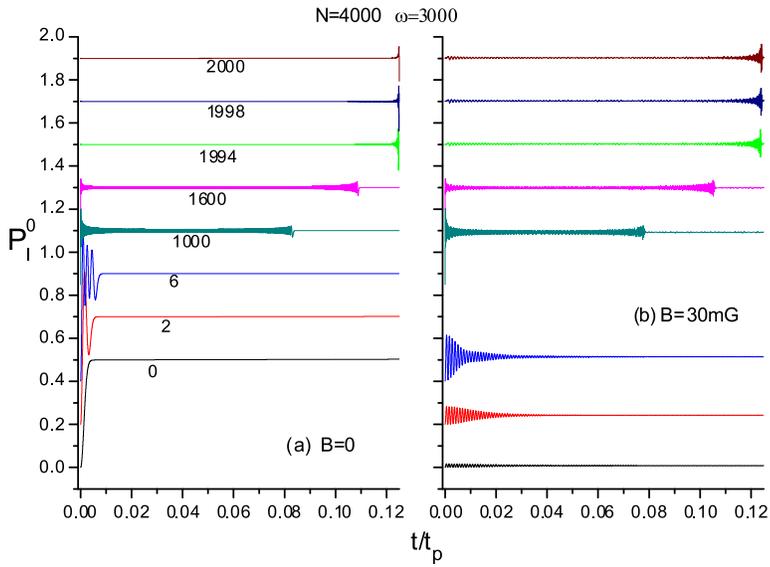}}
\caption{(colored online) Similar to Fig.\protect\ref{CF1} but with
$N=4000$. $t$\ is from 0 to $t_{p}/8$. A new set of $I$ each is\
marked by the associated curve. } \label{CF6}
\end{figure}

When $N$\ changes, the period is also changed due to $t_{p}\propto N^{3/5}$.
When the time is re-scaled, the effect of $N$\ on the evolution is mild if $%
B=0$. This is shown in Fig.\ref{CF6}a where $N=4000$. Fig.\ref{CF6}a is very
similar to Fig.\ref{CF1}a although $N$\ is ten times larger (note that the
time scales of these two figures are different). In particular, the ZOO\ and
ZOS exist, and the rounds of oscillation contained in the ZOO is also $I/2$%
.\ Furthermore, the ZOO of the dark cyan curve in \ref{CF6}a has also the
duration (0,$t_{p}/12$), however 500 rounds of oscillations are included
inside. Since $g\propto N^{-3/5}$, the increase of $N$\ will increase the
ratio $q/g$, therefore will also amplify the effect of $B$. Comparing \ref%
{CF6}b with \ref{CF1}c, the black curve is more severely suppressed,
the red and blue are more severely pulled down.

\section{Summary}

The effect of the magnetic field on the spin-evolution of zero-polarized $%
^{87}$Rb condensates has been studied. The following points are
summarized:

(i) \textit{Periodicity and inherent symmetry}. This important feature of
 $\mathbf{P}_{I}^{M}$, is spoiled by the field.

(ii) \textit{Background. \ }The field pushes all the $\mathbf{B}_{I}^{0}$
from $\approx 1/2$\ towards $I/N$.

(iii) \textit{Zone of steady evolution}. The ZOS is no more highly steady.
When $B$\ is stronger, the division into ZOO and ZOS does not hold.

\begin{figure}[tbp]
\scalebox{1}{\includegraphics{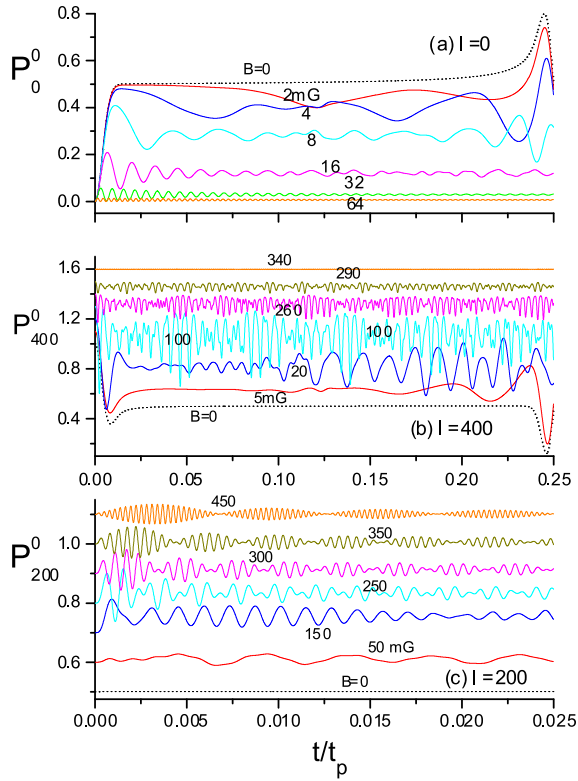}}
\caption{Evolution of $\mathbf{P}_{I}^{0}(t)$ with various $I$\ and
$B$.\ $B$\ is marked by the curves. The dotted lines are
corresponding to $B=0$ . Each curve of (b) and (c)
has been shifted up by 0.1 more than its lower neighbor. $t$\ is from 0 to $%
t_{p}/4$ (a and b) or to $t_{p}/40$ (c). }
\label{CF7}
\end{figure}

(iv) \textit{Sensitivity}. \ When $I$ is close to $0$ or $N$, $\mathbf{P}%
_{I}^{0}$\ is highly sensitive to $B$. \ Otherwise, it is less
sensitive. This is further shown in Fig.\ref{CF7}. \ From this
figure one can see that $B$ causes oscillation in general. The
frequency of oscillation will increase if $B$ increases. The
amplitude will firstly increase with $B$.  However, when $B$ exceeds
certain values, the amplitude decreases with $B$. When $B$ is
sufficiently large, the amplitude will become so small that
$\mathbf{P}_{I}^{0}$ looks just like the horizontal line $I/N$. This
situation would occur if $B > 60,  320$, and 2000 mG, respectively,
when $I=0, N$, and $N/2$. It implies that the suppression of the
amplitude is much easier to be fulfilled if $I \approx 0$ or $N$. On
the other hand, in \ref{CF7}c with $I=N/2$, $\mathbf{P}_{N/2}^{0}$
is only weakly disturbed even if $B$\ is as large as $50$mG (red
curve), and the amplitude is still not very small even if $B$\ is as
large as $450$mG (orange curve).

(v) \textit{Strong oscillation. }When $I$ is close to $N$, a mediate
field (roughly, $30\leq B\leq 180$ mG) causes a particularly strong
oscillation of $\mathbf{P}_{I}^{0}$ through out nearly all the time
as clearly shown in Fig.\ref{CF7}b\ . This is a noticeable point.

(vi) \textit{New modes of oscillation. } Fig.\ref{CF7}c shows clearly how
the increase of $B$\ leads to the appearance of the new mode with olivary
shape of pulses. That is similar to the quantum beats.

\begin{acknowledgments}
We appreciate the support from the NSFC under the grants 10574163, 90306016,
and 10674182.
\end{acknowledgments}

\clearpage

\end{document}